\documentclass{emulateapj}
\usepackage{natbib}
\usepackage{amssymb}
\usepackage{amsmath}
\usepackage{graphicx}
\usepackage{ulem,color}

\def \Ms {M_\odot}

\shorttitle{The Observable Signatures of  GRB Cocoons}
\shortauthors{Nakar  \& Piran}

\begin{document}


\title{The Observable Signatures of GRB Cocoons}


\author{Ehud Nakar\altaffilmark{1}}\email{udini@wise.tau.ac.il}
\author{Tsvi Piran\altaffilmark{2}}


\altaffiltext{1}{The Raymond and Beverly Sackler School of Physics and Astronomy, Tel Aviv University, Tel Aviv 69978, Israel}
\altaffiltext{2}{Racah Institute of Physics, The Hebrew University of Jerusalem, Jerusalem 91904, Israel}


\begin{abstract}
As a long GRB jet  propagates within the surrounding stellar atmosphere it creates a cocoon composed of an outer Newtonian  shocked stellar material and an inner (possibly relativistic) shocked jet material.
The jet deposits $10^{51}-10^{52}$ erg into this cocoon. This energy is comparable to the GRB's energy and to the energy of the accompanying supernova, yet its signature has been largely neglected so far. 
{A fraction of the cocoon energy is released during its expansion following the breakout from the star and later as it interacts with the surrounding matter. We explore here the possible signatures of the cocoon emission and outline a framework to calculate them from the conditions of the cocoon at the time of the jet breakout. We show that the cocoon signature depends strongly on the level of mixing between the shocked jet and shocked stellar material that fills it, which is currently unknown. We find that if there is no mixing at all then the $\gamma$-ray emission from the cocoon is so bright that it should have been already detected, and the lack of such detections indicates that mixing at some level must take place. We calculate also the expected signal for partial and full mixing.}
While the typical signals are weaker than GRBs' afterglows, the latter are highly beamed while the former have  wide angles.  
We predict that future optical, UV and X-ray transient searches, like LSST, ZTF, ULTRASAT, ISS-Lobster  and others will {most likely} detect such signals, providing a wealth of information on the progenitors and jets of GRBs. 
While we focus on long GRBs, we note that analogous (but weaker)  cocoons may arise in short GRBs as well. Their signatures might be the most promising electromagnetic counterparts for gravitational waves merger's signals. 

\end{abstract}



\section{Introduction}

According to the Collapsar model \citep{Woosley93,MacFadyen99}  long GRBs (LGRBs) arise during the collapse of  massive stars. 
Following the collapse, a central engine (either an accreting compact object or a magnetar) is formed  and it drives a bi-polar relativistic jet that  punches a hole through the stellar envelope. The jet produces the prompt GRB and the subsequent afterglow once it is far outside  the star. Association of LGRBs with star forming regions \citep{Paczyski98,Fruchter06} and the observations of associated powerful type Ic supernovae (SNe)   \citep[see e.g.][for a review]{Woosley06} support this model. 

So far these two component: the relativistic jet that produces the observed GRB and its afterglow
and the associated SN, were the focus of attention, both from a theoretical and from an observational points of view. However, the Collapsar model predicts a third component -- the cocoon. This is an inevitable result of the jet propagation within the stellar envelope.  The cocoon has been largely neglected \citep[see however][]{RamirezRuiz02,Lazzati10} in spite of the fact that it  carries a comparable  amount of energy to the GRB jet or to the SN. Our goal, here, is to discuss the remarkable observational implications of this ingredient of the Collapsar model. We describe a general framework for estimating the different components of the cocoon's emission.

As the GRB jet carves its way through the stellar envelope it dissipates its energy in a  double shock (forward-
reverse) structure that forms at its head \citep{Matzner03, Lazzati05, B11}. The hot
head material spills sideways, forming a cocoon that engulfs the jet and collimates it. 
The dissipated energy is significant. As long as the jet is within the stellar envelope it dissipates almost all its energy. As (baryonic) jets' heads  move rather slowly ($\sim 0.3$c ) it takes a few second to cross the star (with $R_* \approx 10^{11}$ cm). This is comparable to the duration of the later prompt  GRB phase. Evidence for that can be seen in the observed duration distribution of LGRBs \citep{Bromberg12}.
One can expect that the jet's luminosity during the propagation phase is comparable to the jet's luminosity observed later during the prompt GRB phase. 
This implies that the energy given to the cocoon is comparable to the GRB's energy, typically of order of $10^{51}-10^{52}$ erg. 

\begin{figure}
  \begin{center}
    \includegraphics[width=85mm]{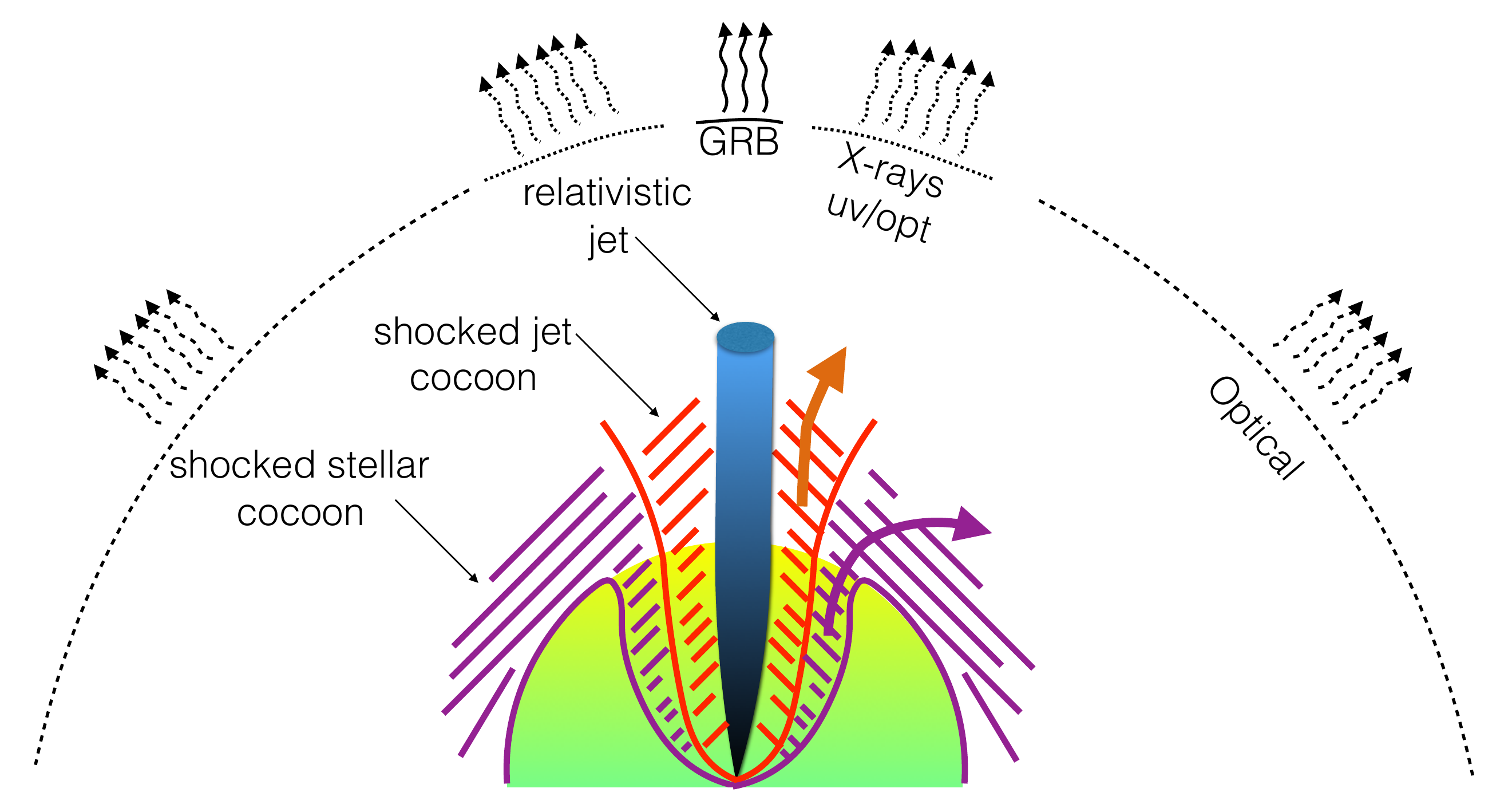}
    \caption{A schematic description of the Collapsar's jets and the cocoon. The cocoon is composed of two components, 
    an inner `shocked jet cocoon' and an outer `shocked stellar cocoon'. The jet cocoon is more dilute and hence it expands to faster, possibly relativistic, velocities.  Also shown are the different emission components and their angular extent. 
    A typical opening angle of the relativistic cocoon components (if exist) is $\sim 0.5$ rad. 
    The stellar  cocoon  is sub-relativistic. As it gets out of the star it engulfs the star and its emission  is practically isotropic. 
}
    \label{fig:schematic}
  \end{center}
\end{figure}

The cocoon is made out of
two components, see Fig. \ref{fig:schematic}, the jet material that was spilled from the head and stellar material that was shocked by the
expanding high pressure cocoon. Numerical simulations of unmagnetized hydrodynamic jets
suggest that while some mixing of the two components may take place, they do remain separated to a large extent (\citealt{Morsony07,Mizuta09,Mizuta13,Lopez13,Lopez16}; Harrison et al., in preparation). The main difference between the two components is that the jet material is much more diluted than the shocked stellar material.  Having the same pressure this
implies that the jet material has more energy per baryon.  As a result it accelerates after thebreakout to higher (possibly relativistic) velocities.
The evolution may be different if the jet is Poynting flux dominated \citep{Levinson13,Bromberg14}.  It depends strongly on the stability of the jet and the location where the magnetic field is dissipated. \cite{Bromberg16} find that it is dissipated early on and  the larger scale evolution will resemble a baryonic jet. The main difference is that in this case the shocked cocoon material will be magnetized while the stellar cocoon won't, and it is likely that mixing will be somewhat suppressed as compared with the mixing in a purely baryonic jet.

The cocoon emission, from each one of the two components has two sources: (i) Diffusion of
the internal energy deposited during the jet propagation.
This emission is similar to the cooling envelope emission
of a SN, or to the so called photospheric emission of a GRB fireball. Here we denote this phase as the `cooling cocoon
emission'. (ii) Interaction of the cocoon material with the
external medium. This emission is particularly important for the jet cocoon, which could be relativistic. In  this case the emission is 
similar in nature to the GRB afterglow. Here we denote this phase as the
`cocoon afterglow'. 

Already in 2002, \citet{RamirezRuiz02} realized the importance of the cocoon component and discussed its possible signature. However, these authors considered only the shocked jet component assuming that it has the same Lorentz factor as the GRB jet, ignoring the important possibility  of mixing between the shocked stellar material and the shocked jet material. This assumption implies a relativistic expansion of the shocked jet, like a fireball. 
They realize that the emerging flow would result in a wider angle than the GRB jet and that this may lead to an off-axis   `cocoon afterglow' that can be observed also in cases that the $\gamma$-rays are not seen. \citet{RamirezRuiz02} also discuss a possible photospheric 
 `cooling emission' 
from this fireball.  
\cite{Lazzati10}  carried out a numerical simution of cocoon evolution. Using the energy distribution per solid angle and an ad hoc emission model they estimated the resulting prompt $\gamma$-ray emission that  will be seen by an observer at 45$^o$ finding that it may resemble a short GRB. 

We present here an analytic framework for estimating the different cocoon emission components, 
both from the  shocked jet and from the shocked stellar material, that has been ignored so far. 
In this framework we estimate the cocoon's signature as a function of the cocoon's parameters at the time 
that it breaks out from the stellar envelope. 
{The signature of the shocked jet depends strongly on the amount of  mixing between the two cocoon's components, which is currently unknown. This amount is likely to depend on the properties of the jet and the progenitor, and it should be explored by detailed numerical simulations. We therefore leave it as a free parameter and calculate the resulting emission as a function of the mixing.   
We focus, however, on the signatures resulting from partial mixing as suggested by previous numerical simulations \citep[e.g.,][]{Morsony07,Mizuta09,Mizuta13,Lopez13,Lopez16} and by our preliminary numerical results  (Harrison et al., in preparation).}

{While we focus on cocoons arising in LGRBs as suggested by  the Collapsar model, we note that cocoons are also expected in short GRBs, if those are generated by merger of two neutron stars, and that our formalism  
can be applied to them as well. When two neutron stars merge, matter is ejected prior to the jet's onset by tidal forces, by winds  driven from the newly formed hypermassive neutron star and from the debris disk that forms around it \citep[see e.g.][for a brief review]{Hotokezaka15}.   
The cocoons are generated during the interaction of the GRB jet with this surrounding matter \citep{Nagakura14,MurguiaBerthier16}. 
These cocoons will be much less energetic than those produced in LGRBs, reflecting the fact that LGRBs
are much more energetic than short GRBs. Still they may lead to a detectable signal from events taking place at
a few hundred Mpc from us, giving rise to a new potential EM counterpart to the gravitational radiation signals arising from these mergers. } 

Cocoon emission would arise also in failed LGRBs where the jets are chocked before they break out. 
This happens when the central engine stops early enough before the jet reaches the outer edge of the stellar envelope, so the entire launched jet ends up in the cocoon.  
Chocked jets may be quite numerous, in fact the duration distribution of GRBs suggests that they are much more numerous  then successful GRBs \citep{Bromberg12}. 
When the jet is choked it dissipates all its energy within the stellar envelope and  there is no GRB. However if before chocking the jet  crossed a significant fraction of the stellar envelope then the cocoon is energetic enough to break out of the star by itself and produce an observable signature. 

Various authors \citep{Kulkarni98,Tan01,Macfadyen01,Campana06,Wang07,Katz10,NS10,Bromberg11a}  suggested  that the emission from the shock breakout of chocked jets'  cocoons produce
the  low luminosity GRBs ({\it ll}GRBs).  However \cite{NS12} have shown that shock breakout can produce the observed {\it ll}GRBs only if their progenitors are extended ($>10^{12}$ cm). In particular \cite{Nakar15} have shown that both the signature of GRB 060218 and the accompanying SN2006aj show that indeed the progenitor star must have had a low mass extended envelope of $\sim 10^{13}$cm.  So, if  {\it ll}GRBs do arise form choked jets they must involve altogether a different population of progenitors than regular LGRB progenitors. 
In this paper we focus on regular LGRBs and do not explore {\it ll}GRBs.
 
We begin in \S \ref{sec:evolution}
 with a discussion of  the properties
of the shocked stellar cocoon and the shocked jet cocoon at the time of the jet breakout. We continue  in \S \ref{sec:dynamics} describing  the subsequent dynamics after breakout.
We turn in \S \ref{sec:emission} to the emission from the different components. In 
\S \ref{sec:detectability} we discuss the detectability of these signatures using current and future telescopes.  
In \S \ref{sec:SGRBs} we discuss possible cocoon signature in short GRBs, before concluding in \S \ref{sec:conclusions}.


\section{The cocoon's energy, volume and mass at  breakout}
\label{sec:evolution}

As the jet breaks out of the star it expands  and its energy is not dissipated anymore. While the cocoon continues to collimate the jet for some time, this does not strongly affect the 
the total cocoon's energy, which remains rather constant. After breakout the cocoon is also free to expand.  The lighter  shocked jet material expands first, followed by the slower Newtonian shocked envelope material. The properties of the expanding cocoon material, and thus its emission, depend on the conditions at the time of breakout, which we discuss below.

The most important properties of the cocoon, for our purpose, are its energy, size and mass density profile at the time of breakout\footnote{Another important property
is the opacity, which in turn depends on the  composition. While simple for LGRB cocoons, this brings another source of uncertainty for short GRB cocoons that we discuss later. }. The total cocoon energy, $E_c$, is expected to be comparable to the total GRB energy (prompt emission and afterglow kinetic energy). The reason is that the typical breakout time is comparable to a typical burst duration \citep{Bromberg12} and the jet deposits almost all its energy into the cocoon during its propagation in the progenitor.  The GRB's energy is the jet's energy after the breakout and there is no reason to expect that the jet won't have the same luminosity before and after breakout. The total energy distribution of GRBs spans over several decades, centering between $10^{51}$ erg and $10^{52}$ erg \citep[e.g.,][]{Cenko10,Shivvers11}. We use, therefore,  $E_{51.5}=E_c/10^{51.5}$ as the canonical value for the cocoon's energy. 

If the dependence of the jet breakout time on the jet's and progenitor's properties is known, then the cocoon energy can be directly related to them. In the case of a purely hydrodynamic jet  we can use the analytic modeling of \cite{B11} calibrated by numerical simulations. This model is also expected to be applicable to magnetized jets since numerical simulations indicate that they dissipate a large fraction ($\sim$ half) of their magnetic energy deep within the progenitor due to instabilities \citep{Bromberg16}, propagating the rest of the way similarly to hydrodynamic jets.     
The total energy deposited by the jet in the cocoon until the breakout time, $t_b$, is  $E_c=\int_0^{t_b} L_j (1-\beta_h) dt$, where $L_j$ is the total (two-sided\footnote{Note that in \citealt{B11} the jet luminosity considered is only that of one side of the bi-polar jet})  jet luminosity,  $\beta_h c$ is the velocity of the Jet's head and $c$ is the light speed. For typical jet and stellar parameters the jet propagates in the star at Newtonian to mildly relativistic velocities. Therefor we can approximate $E_c \approx L_j t_b$. The time to breakout is:
\begin{equation}\label{eq:tb}
	t_b \approx 8  ~L_{51}^{-1/3} \theta_{10^o}^{4/3} R_{11}^{2/3} M_{10}^{1/3}~{\rm ~s},
\end{equation}
where $L_{51}=L_j/(10^{51} ~\rm{erg/s})$, $\theta_{10^o}$ is the jet's half opening angle, $\theta_j$, in units of $10^o$, $R_{11}$ is the progenitor's radius in units of $10^{11}$cm and $M_{10}$ is its mass in units of $10M_\odot$.
Here we used the analytic dependence of $t_b$ on the jet and star parameters as found by \cite{B11}, calibrated    
using the numerical result of \cite{Mizuta13}. Thus, the relation between the total energy of the cocoon upon its breakout and the jet and progenitor parameters is:
\begin{equation}
	E_c \approx 8 \times 10^{51}  ~ L_{51}^{2/3} \theta_{10^o}^{4/3} R_{11}^{2/3} M_{10}^{1/3} ~ {\rm ~erg}.  
\end{equation}

The cocoon shape at the time of breakout, as seen in numerical simulations \citep{Morsony07,Mizuta13} and predicted by analytic modeling \citep{B11}, is roughly a cone or a barrel with a hight $R_*$ and a width $\sim  R_* \theta_j$. The total cocoon volume is $V_c \approx \pi R^{3} \theta_j^2$ and the shocked stellar mass in the cocoon is roughly
\begin{equation}\label{eq:mcs}
	m_{c,s} \approx 0.15 \Ms ~ \theta_{10^o}^{2} M_{10} \ .
\end{equation}

\begin{figure}[!t]
\epsscale{1}
\includegraphics[width=65mm,angle=90.]{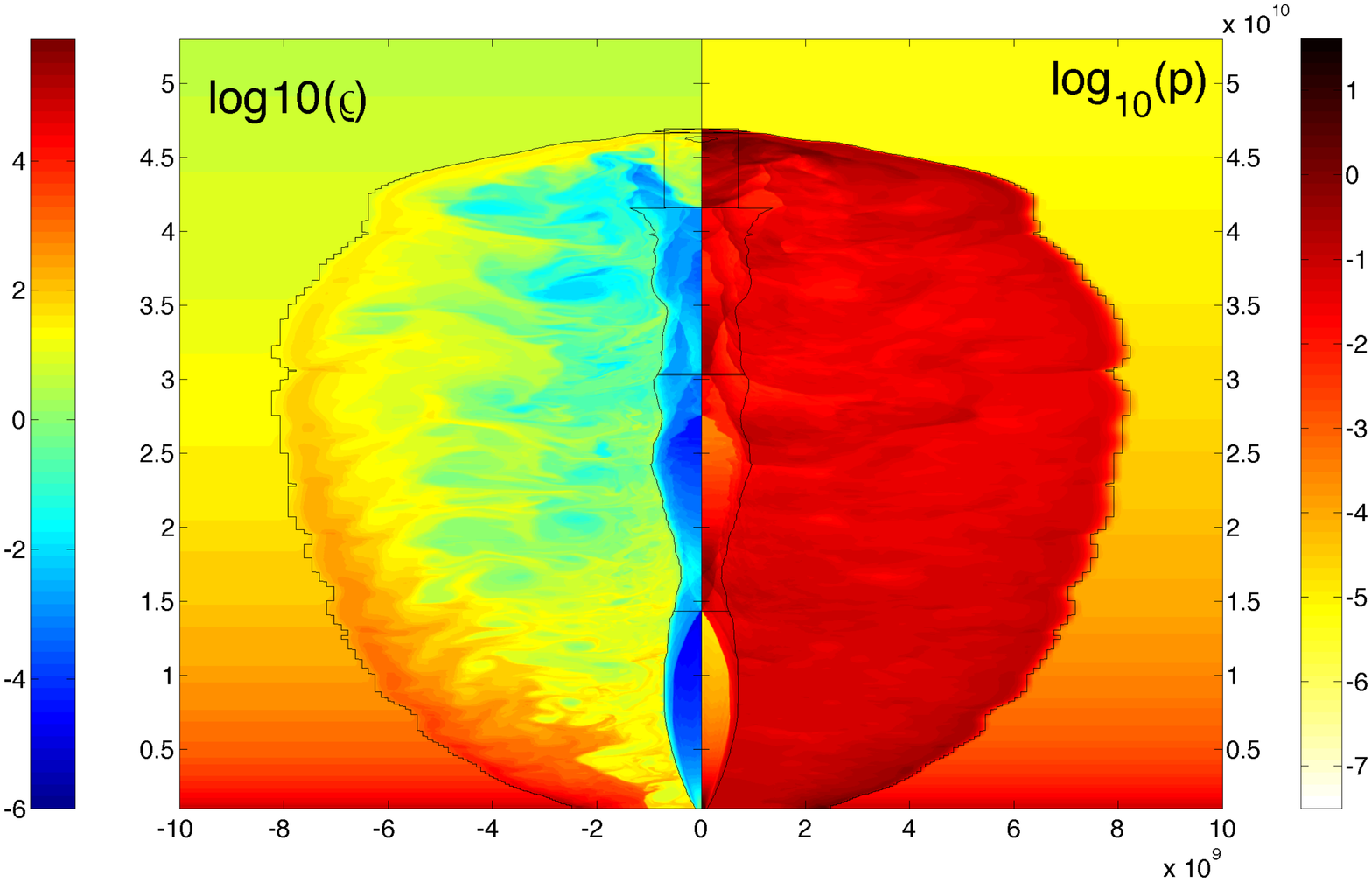}
\caption{The pressure (right) and density (left) within a cocoon arising from a jet propagating {within  external media with a density gradient $\propto r^{-2.5}$.} From a 2D numerical simulation (Harrison et al., in preparation).
One can clearly see that the pressure is more or less uniform, but there is a jump in the density between the cocoons formed out of the shocked jet and the shocked stellar material.}  
\label{fig:p_rho}
\end{figure}

\section{The cocoon dynamics following breakout}
\label{sec:dynamics}
Both the shocked stellar cocoon and the shocked jet cocoon expand once the jet breaks out of the star.
The pressure within the shocked jet and the shocked stellar material
is roughly constant and the main difference between the two regions is the density. This can be seen clearly figure \ref{fig:p_rho} that shows the pressure and density from a 2D numerical simulation\footnote{The simulations we use here were preformed using the public code PLUTO \citep{Mignone07} and are taken from Harrison et al. (in preparation).} of a hydrodynamic relativistic jet that propagates in an external medium.  The energy per baryon varies between the different regions in the cocoon. Following the expansion (almost) the entire energy is carried by bulk motions. Therefore, the energy per baryon of a fluid element in the cocoon provides a reasonable estimate of its velocity after expansion. The shocked jet material will expand first possibly reaching relativistic velocity, while the shocked stellar material will expand more slowly at Newtonian velocities. Below we discuss the dynamics of these two components.

\subsection{The shocked stellar material}
The energy per baryon of the shocked stellar material, and thus its terminal velocity, can be estimated simply using the  deposited energy, $\approx E_c/2$,  and the mass, $m_{c,s}$. The terminal velocity of the shocked stellar material after the expansion is 
\begin{equation}\label{eq:vcs}
	v_{c,s} \approx \sqrt{\frac{E_c}{m_{c,s}}} \approx 3 \times 10^{9}  ~E_{51.5}^{1/2} \theta_{10^o}^{-1} M_{10}^{-1/2} {\rm ~cm/s} \ .
\end{equation}  
At this velocity the shocked stellar material spills out of the cocoon's opening and it spreads sideways almost spherically. The optical depth at the time of breakout is $\gg c/v$ and almost the entire internal radiation energy is converted to the kinetic energy of the outflow. By the time that the shocked stellar material  expands to $\sim 2 R_*$  it has practically  accelerated to its terminal velocity and it approaches a spherical homologous expansion. A small fraction of the initial energy will be later released {once the diffusion becomes comparable to the dynamical time.}  

\subsection{The shocked jet material}\label{sec:jetDynamics}
The  shocked jet material dynamics is more complicated.  More importantly it strongly depends on the unknown amount of mixing between the stellar and jet material, that  determines the energy per baryon. We  calculate, therefore, the dynamics as a function of the energy per baryon in the expanding material, using the usual notation of baryon loading
\begin{equation}
	\eta_{c,j} = \frac{E_{c,j}}{m_{c,j}c^2} \ ,
\end{equation} 
where $E_{c,j} \approx E_c/2$ is the energy deposited in the cocoon shocked jet material and $m_{c,j}$ is its mass. If $\eta_{c,j} \lesssim 1$, the jet material expands roughly spherically to a terminal velocity of about 
\begin{equation}
	v_{c,j} \approx \sqrt{\frac{E_c}{m_{c,j}}} ~~~;~~~ \eta_{c,j} \lesssim 1 .
\end{equation}
If, however, $\eta_{c,j} \gg 1$ the shocked jet material  reaches  a relativistic velocity. The dynamics of the acceleration is similar to that of a baryon loaded relativistic fireball \citep{RamirezRuiz02}. The dynamics  of relativistic fireballs has been  discuss by many authors \citep[see e.g.][]{Goodman86,Shemi90,Meszaros93,Piran93,Grimsrud98,Daigne02,Nakar05}\footnote{Some of the early fireball studies had various errors, see  \cite{Nakar05} for a discussion.}.  We follow here \cite{Nakar05}.

The scenario we consider is that of a hot and dilute medium, initially at rest, with a constant $\eta_{c,j} \gg 1$ that fills a roughly cylindrical\footnote{The differences between a cylindrical and a conical cavity results in  factors  of order unity.} cavity with a hight $\sim R_*$ and a radius $\sim R_* \theta_j$. The cavity is open in one of the cylinder bases and the hot material is free to expand out of the cylinder where the density is negligible. As soon as the hot material gets out of the cavity it expands sideways and accelerates simultaneously. The sideway expansion stops once the opening angle is $\sim 1/\Gamma$, where $\Gamma$ is the  instantaneous Lorentz factor of the outflow. From this point onwards the outflow expands conically following the regular fireball theory. Since the initial velocity of the outflow is not relativistic its final opening angle is relatively large, $\theta_{c,j} \sim 0.5$ rad.

In the regular fireball theory the outflow is spherical. Its evolution depends on the initial radius, the outflow luminosity and its baryon loading. In our case the initial radius is the size of the cavity opening, $\sim R_* \theta_j$, and the luminosity is $\sim E_{c,j} c/R_*$. In addition, since the the evolution depends on the isotropic equivalent luminosity, the actual outflow luminosity should be divided by $\theta_{c,j}^2/2$, the fraction of the solid angle that the outflow covers out of the entire $4\pi$ sphere. The outflow's evolution depends on the relation between the actual baryonic loading and the critical baryonic load\footnote{The critical baryonic loading denoted here $\eta_b$ is denoted as $\eta_3$ in  \cite{Nakar05}.} \citep{Nakar05}:
\begin{equation}
	\eta_b = \left( \frac{E_{c,j} \sigma_T}{2\pi \theta_{c,j}^2 R_*^2 \theta_j m_p c^2}  \right)^\frac{1}{4} \approx 125 E_{51.5}^{1/4} \theta_{10^o}^{-1/4} R_{11}^{-1/2} \theta_{c,j,0.5}^{-1/2}  \ ,  
\end{equation}
where $\sigma_T$ is the Thomson cross-section and $m_p$ is the proton mass\footnote{This equation corrects  Eq. 6 of \citet{RamirezRuiz02} that has a different power (1/3 instead of 1/4 here) arising from an earlier wrong model of a fireball evolution and has a wrong  power of $\theta_j$ in the denominator. As as result of these differences our value of $\eta_b$ is much lower. This has significant observational implications.}. 
For  $\eta_{c,j} < \eta_b$ almost the entire radiation energy is converted to the bulk kinetic energy  of the baryons, that are accelerated to a terminal Lorentz factor $\Gamma_{c,j}=\eta_{c,j}$. In addition, the radiation remains trapped also after acceleration ends and it suffers adiabatic loses before being released to the observer. If, however,  $\eta_{c,j} > \eta_b$ the radiation is released before acceleration ends and it carries most of the initial energy. The terminal Lorentz factor of the outflow in this case is $\eta_b$. Therefore, the terminal Lorentz factor of the shocked jet material is:
\begin{equation}
	\Gamma_{c,j}= \min\{\eta_{c,j},\eta_b\}~~~;~~~ \eta_{c,j} \gg 1 \ .
\end{equation}
The terminal kinetic energy of the outflow is
\begin{equation}
	E_{k,cj}=E_{c,j} \cdot \min \left\{ 1,\frac{\eta_b}{\eta_{c,j}} \right\} \ . 
\end{equation}

Our analytic modeling of the cocoon does not provide the actual value of $\eta_{c,j}$. Moreover, it is possible, and even likely, that this value depends on the properties of the jet or those of the  stellar atmosphere (e.g., magnetic fields tend to suppress mixing). Therefore, we  consider several constraining limits. In the limit of a full mixing $m_{c,j} \approx m_{c,s}$ and the shocked jet component is similar to the shocked stellar material discussed above. In the other limit of no mixing $\eta_{c,j} = \Gamma_j$, the jet's Lorentz factor. Since compactness limits indicates that typically $\Gamma_j \gtrsim 100$, then if there is no mixing $\eta_{c,j} \gtrsim \eta_b$ and the escaping radiation carries an energy of $\sim E_{c,j}$ while the kinetic energy of the outflow is comparable or smaller. As we see later (section \ref{sec:JetCollingEmission}) observations seems to rule out this possibility.

\begin{figure}[!t]
\epsscale{1}
\includegraphics[width=65mm,angle=90.]{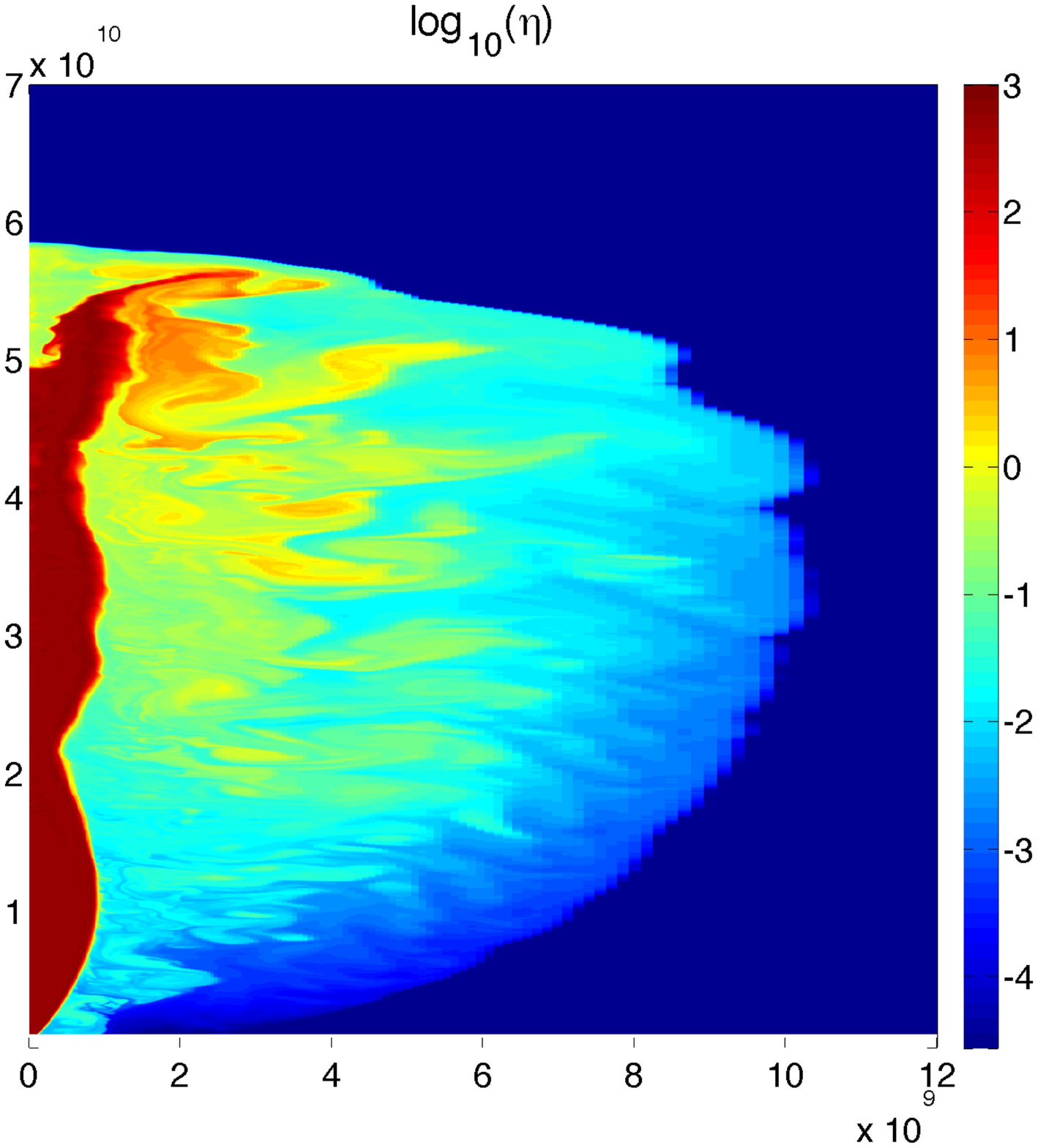}
\caption{ The local value of the baryonic loading, $\eta$ in logarithmic scale from a 2D simulation of a hydrodynamic jet propagation in a stellar atmosphere with a density 
gradient $\propto r^{-2.5}$ (same as figure \ref{fig:p_rho}; from Harrison et al., in preparation).  The shocked jet and the shocked stellar material remains largely  separated, namely there is no full mixing.  The shocked jet material in the cocoon has a range of baryonic loading values, $\eta_{c,j} \sim 0.01-10$, compared to the the baryonic loading of the jet itself which is $\eta=550$, indicating  a significant yet partial mixing. The figure also shows that the higher $\eta$ (lighter) material is on top of lower $\eta$ (heavier) material. This implies that after the breakout the lighter material is free to expand first to high velocities and evacuate the cocoon cavity, while the slower material is free to expand later to slower velocities. Thus, the value of the baryonic loading of a fluid element in the shocked jet cocoon at the time of breakout provides a reasonable approximation to its value after expansion. }   
\label{fig:eta}
\end{figure}

Between these two limits there is partial mixing:  the shocked jet material expands faster than the shocked stellar material but at a lower velocity than  the jet (although it may still be relativistic).
In this case the kinetic energy of the outflow  carries most of the initial energy and the escaping radiation is suppressed by adiabatic cooling (see section \ref{sec:JetCollingEmission}). Numerical simulations of (unmagnetized) 
hydrodynamic jets suggest that this is the relevant case \citep[e.g.,][]{Lazzati10,Mizuta13,Lopez16}.  Figure \ref{fig:eta} illustrates an example that 
shows the result of a 2D simulation in which a hydrodynamic jet propagates in external medium with a density 
gradient $\propto r^{-2.5}$ (taken from Harrison et al., in preparation). It shows the local value of $\eta$ in a logarithmic color scheme. One can see that the shocked jet and the shocked stellar material remains largely 
separated, namely there is no full mixing. Instead, jet material goes through different levels of mixing in 
the range  $\eta \sim 0.01 - 10$, where material closer to the head of the jet has higher $\eta$ value (i.e. it is less mixed) than material near the base of the jet. 
The results of this specific numerical simulation suggest that the mixing is such  that after expansion roughly an equal amount of energy is carried out per every logarithmic scale of $\Gamma\beta$ in the range $0.1-10$. 

\section{Emission}
\label{sec:emission}
We turn now to consider the emission from the different shocked components. We begin with the shocked stellar material and then turn to the shocked jet. 

\subsection{The emission from the shocked stellar material}
\label{sec:stellar}
The physics of the emission from the expanding shocked stellar material is similar to that of a cooling envelope in a supernova. {As long as the gas is ionized and the expansion approximated as spherical, the main properties of the emission can be approximated using simple arguments \citep[e.g.,][]{Arnett80,Kasen09,NakarPiro14}. Here we briefly repeat this derivation and then apply it to our case.}

Consider a spherical shell with a mass, $m$, and an opacity per unit of mass, $\kappa$, that expands homologously at a characteristic velocity, $v$. The internal radiation in such a shell diffuses to the observer once its optical depth, $\tau \approx \kappa m/(4 \pi v^2 t^2)$, is roughly $c/v$. Namely, at
\begin{equation}\label{eq:tobs}
t_{obs} \approx 0.8 ~\left(\frac{m}{0.01\Ms}\right)^{1/2}\left(\frac{v}{10^{9} {\rm ~cm/s}}\right)^{-1/2} \kappa_{0.1}^{1/2} {\rm ~d} \ .
\end{equation}
where $\kappa_{x}=\kappa/(x {\rm \frac{cm^2}{gr}})$. An internal energy, $E_{r,0}$, was deposited initially in this shell  in the form of radiation when its mass occupied a volume $V$.  By the time  $t_{obs}$ the radiation has cooled adiabatically and its energy is roughly $E_r(t_{obs}) \sim E_{r,0} V^{1/3}/(vt_{obs})$.  Since this energy is released over $t_{obs}$ the observed luminosity is $\sim E_r(t_{obs})/t_{obs}$. In our case the cocoon occupies at breakout a volume of $\sim \pi R_*^3 \theta_j^2$ and its internal energy is   $E_{r,0} \sim mv^2/2$. This implies a luminosity:
\begin{equation}\label{eq:Lobs}
	L(t_{obs}) \approx 5 \times 10^{40} ~R_{11} \theta_{10^o}^{2/3} \left(\frac{v}{10^{9} {\rm ~cm/s}}\right)^{2} \kappa_{0.1}^{-1}~{\rm erg/s} \ .
\end{equation}
The color temperature at $t_{obs}$ can be approximated by the effective temperature
\begin{equation}\label{eq:Tobs}
	T(t_{obs}) \approx 10^4~\left(\frac{t_{obs}}{day}\right)^{-1/2} R_{11}^{1/4}\theta_{10^o}^{1/6} \kappa_{0.1}^{-1/4} {\rm~K} \ .
\end{equation}

The opacity, $\kappa$,  depends on the gas temperature. The cocoon is expected to be dominated by C and O (GRB progenitors show no evidence of H and He, while Ni is not expected to be mixed into the upper layers of the cocoon). The typical gas density at the time that the luminosity peaks is $\lesssim 10^{-11} {\rm ~gr/cm^3}$, where the Rosseland  opacity\footnote{Opacities are taken from http://opacities.osc.edu/rmos.shtml \citep{Seaton05}} of a C/O mixture ranges from $0.03 {\rm ~cm^2/gr}$ at $\sim 7000$ K to $0.2 {\rm~cm^2/gr}$ at 30,000 K. Below 7000 K the gas recombines and the number of free electrons drop sharply, and so does the  opacity. Note that the opacity that determines the observed luminosity is set by gas that is at an optical depth $\sim c/v$  where the temperature is larger by  a factor of about $(c/v)^{1/4}$ than the effective temperature.

The cooling emission depends on the mass, energy, initial radius and opening angle of the outflow. All these quantities, and their dependence on the progenitor and jet properties, are rather well known for the shocked stellar material and therefore the predicted signal of this component is robust.
Plugging $m_{c,s}$ and $v_{c,s}$ (Eqs. \ref{eq:mcs} \& \ref{eq:vcs})  into eqs. \ref{eq:tobs} - \ref{eq:Tobs} we find that the cooling emission from the shocked stelar material peaks at
\begin{equation}\label{eq:tcs}
	t_{c,s} \approx 1.2 ~E_{51.5}^{-1/4} \theta_{10^o}^{3/2} M_{10}^{3/4} \kappa_{0.05}^{1/2} {\rm ~d}
\end{equation}
after the GRB at a luminosity
\begin{equation}\label{eq:Lcs}
	L_{c,s} \approx 10^{42} E_{51.5} \theta_{10^o}^{-4/3} R_{11} M_{10}^{-1} \kappa_{0.05}^{-1}~{\rm erg/s} \ .
\end{equation}
The color temperature at $t_{obs}$ can be approximated by the effective temperature
\begin{equation}\label{eq:Tcs}
	T_{c,s} \approx 9,000~ E_{51.5}^{1/8} \theta_{10^o}^{-7/12} R_{11}^{1/4} M_{10}^{-3/8} \kappa_{0.05}^{-1/2} {\rm~K} \ .
\end{equation}
Here we took a canonical $\kappa=0.05{\rm ~cm^2/gr}$, which is appropriate for a C/O gas at $10^4$ K. 

We find that a the cocoon emission from a typical GRB will produce an isotropic optical signal a day after the event. The signal is rather bright at an absolute magnitude $\sim -16$ and is dominant over the  ``orphan afterglow", namely an afterglow from a  typical GRB  seen off-axis at a large viewing angle  \cite[e.g.][]{Nakar02}, at the same time (the latter may become brighter later and overshine this signal).

\subsection{Emission from the shocked jet}
\label{sec:emission-jet}
As discussed above, the emission from the shocked jet depends on the mixing. Since the mixing level is unknown we consider three option, no mixing, full mixing and partial mixing. 
If there is full mixing then there is no difference between the jet and the stellar material. The emission is similar to the one discussed in the previous section. We consider this case as unlikely since it is not supported by numerical simulations. In the cases of no mixing, or partial mixing the jet shocked material has a distinct observational signature that we discuss below. We separate the discussion to the roughly isotropic emission from the cooling emission of material that expands in Newtonian velocities, the cooling emission of material that expands relativistically and the afterglow emission generated by interaction of therelativistic cocoon outflow with the external medium.

\subsubsection{Isotropic cooling emission from a Newtonian jet material ($\eta_{c,j} \lesssim 1$)}
A distinctive signature from Newtonian shocked jet material is expected only if there is a partial mixing (there is no Newtonian shocked jet component if there is no mixing at all). The emission from such a  component is similar to that of the shocked stellar material, but being faster and lighter it is brighter, it peaks earlier and at shorter wavelengths. A significant contribution from this component (compared to the shocked stellar material) is expected only if non negligible fraction of the cocoon energy is deposited in material with $\beta\Gamma  \approx 1$. We  parametrized this unknown fraction as $f_{\beta \Gamma,1}$.  Numerical simulations of hydrodynamic jets suggest that $f_{\beta \Gamma,1} \sim 0.1$, and  we will use this as a canonical value. This mildly relativistic material will also generate the peak of the jet component isotropic emission.

Using equations \ref{eq:tobs}-\ref{eq:Tobs} with $\beta=0.7$ we obtain that the peak of the  Newtonian jet component  signal is observed at:
\begin{equation}\label{eq:tcjN}
	t_{cj,N}^{peak} \approx 1.6 ~E_{51.5}^{1/2} \kappa_{0.2}^{1/2} \left( \frac{f_{\beta \Gamma 1}}{0.1}\right)^{1/2}{\rm ~hr} \ , 
\end{equation}
after the GRB. The peak  luminosity and the observed temperature are:
\begin{equation}\label{eq:LcjN}
	L_{cj,N}^{peak} \approx 10^{43} \theta_{10^o}^{2/3} R_{11} \kappa_{0.2}^{-1}~{\rm erg/s} \ ,
\end{equation}
and
\begin{equation}\label{eq:TcjN}
	T_{cj,N}^{peak} \approx 28,000~E_{51.5}^{-1/4} \theta_{10^o}^{-1/6} R_{11}^{1/4} \kappa_{0.2}^{-1/2} \left( \frac{f_{\beta \Gamma 1}}{0.1}\right)^{-1/4} {\rm~K} \ ,
\end{equation}
where we use $\kappa_{0.2}$ as the canonical opacity as appropriate for the expected temperature. Note that this component is isotropic. 

The emission following the peak depends on the energy distribution as a function of the velocity. We will parameterize this distribution as a power-law $dE/dv \propto v^{-s}$, assuming $s>-1$. Numerical simulations suggest roughly a constant amount of energy per logarithmic velocity scale, namely $s \approx 1$. As $dE/dv \propto m v$ we obtain $m(>v) \propto v^{-(s+1)}$ (the assumption $s>-1$ implies more mass is moving at lower velocity than at higher velocity). For the expected observed density and temperature range the opacity depends on the temperature. A rough approximation at the range 7,000-30,000 K is $\kappa \propto T^{1.3}$. Plugging these relations to equation \ref{eq:tobs}-\ref{eq:Tobs} we obtain 
\begin{equation}\label{eq:LcjN_evolution}
	L_{cj,N} \propto t^{-\frac{4+0.5s}{2+s}} 
\end{equation}
and
\begin{equation}
\label{eq:TcjN_evolution}
	T_{cj,N} \propto t^{-0.38} .
\end{equation}
The temperature evolution is independent of $s$ while for $s=1$ the luminosity evolves as $L_{cj,N} \propto t^{-1.17}$. Note that the Newtonian material achieves a thermal equilibrium while the cocoon is still trapped, before it breaks out (see next subsection). Therefore the derivation above, which assumes thermal equilibrium is applicable.  

To conclude, for our canonical parameters the isotropic signal from the cocoon shocked jet material peaks about an hour after the GRB with a very blue UV-optical spectrum ($\sim 30,000$ K). The absolute magnitude in the near-UV is $\sim -18$ while in the optical blue bands it is $\sim -17$. With time both the luminosity and the temperature drop leading to a slow decline in the observed optical/UV signal. The optical emission from the shocked stellar material is expected to peak after $\sim 1$ d at a similar, or slightly fainter, magnitude compared to that of the shocked jet optical peak. In UV, however,  the shocked stellar material emission is much fainter than the UV peak of the shocked jet material. 

\subsubsection{Photospheric (cooling) emission from a relativistic material ($\eta_{c,j} \gg 1$)}\label{sec:JetCollingEmission}
Here we calculate the emission from a relativistic shocked jet material. This include the case of no mixing, where the shocked jet material has a similar baryonic loading as the jet itself, i.e., $\eta_{c,j}=\Gamma_j$. It also includes cases of partial mixing where parts (or all) of the shocked jet material is accelerated to, possibly a range of, relativistic velocities. Cocoon material with $\eta_{c,j} \gg 1$ expands like a fireball with a wide opening angle, $\theta_{c,j} \approx 0.5$ rad, as discussed in section \ref{sec:jetDynamics}. The fireball evolution, and the observed emission, depend strongly on the relation of between $\eta_{c,j}$ and $\eta_b$. As it turns out in case of no mixing $\eta_{c,j}=\Gamma_j \gtrsim \eta_b$. Therefore, the difference between no mixing and partial, yet significant, mixing is dominated by the fact that in the former  $\eta_{c,j} \gtrsim \eta_b$ while in the latter $\eta_{c,j} < \eta_b$. Below we discuss each of these cases separately. To calculate the emission in each case we first find the temperature of  the radiation at the time of the breakout and then for each case we follow the radiation temperature and energy up to the photosphere where it is release.

At the time of the breakout the entire cocoon energy is deposited in radiation that occupies a volume $V_c \sim \pi R_*^3 \theta_j^2$. Therefore, if the radiation is in thermal equilibrium its initial temperature at the beginning of the fireball acceleration is:
\begin{equation}
	T_{BB,0} = \left( \frac{E_c}{V_c a_{BB}} \right)^\frac{1}{4} \approx  
	20 ~E_{51.5}^{1/4} \theta_{10^o}^{-1/2} R_{11}^{-3/4} {\rm~keV} ,
\end{equation}   
where $a_{BB}$ is the radiation constant. However, the radiation is not necessarily in thermal equilibrium. To reach and maintain thermal equilibrium the gas must produce enough photons in the available time (see \citealt{NS10} for a detailed discussion). The main photon generation process at these temperatures where the gas is fully ionized is free-free and therefore we can use the criterion for thermal equilibrium derived in \cite{NS10}. Taking the available time for thermalization as $\sim t_b$, the mass density as $E_{c,j}/(\eta c^2 V_{c,j})$ and the temperature as $T_{BB,0}$ and plugging these values to equation\footnote{Note that the notation $\eta$  in \cite{NS10} refer to the thermalization parameter and is different than the notation $\eta$ used here,  the baryonic loading parameter} 9 in \cite{NS10} we find that the criterion for thermal equilibrium in the cocoon is:
\begin{equation}\label{eq:etaBB}
	\eta_{c,j} \lesssim \eta_{BB}= 50~E_{51.5}^{9/16} \theta_{10^o}^{-9/8} R_{11}^{-27/16} \left( \frac{t_b}{10 {\rm~s}} \right)^{1/2} \ .
\end{equation}
If $\eta_{c,j}>\eta_{BB}$  there is not enough  time to generate enough photons to achieve thermal equilibrium and the initial temperature in the cocoon is higher than $T_{BB,0}$. But, if the temperature is higher than about $50$ keV a significant number of pairs is produced and this significantly increases the photon production rate. Since the number of pairs is exponential with the temperature, pair production behaves as a thermostat that prevents the temperature from rising above  $\sim 100$ keV (for an analog in the structure of relativistic radiation mediated shocks see \citealt{Budnik10,NS12}). \\

\noindent {\it i. No mixing ($\eta_{c,j}>\eta_b$)} \\
The evolution in this case is simple. First,  for reasonable parameters $\eta_{c,j} \gtrsim \eta_{BB}$, implying that the initial temperature of the shocked jet material is $\sim 100$ keV. Second, $\eta_{c,j} \gtrsim \eta_{b}$, implying that the radiation is released during acceleration and therefore it carries almost the entire shocked jet material energy to the observer at a temperature that is comparable to the initial temperature. The duration of the emission is similar to duration over which energy is injected into the fireball:
\begin{equation}
	t_{c,j} \approx \frac{R_*}{c} = 3.3 ~R_{11} ~{\rm s} ~~~;~~~{\rm no~mixing} \ .
\end{equation}
The {\it isotropic equivalent} luminosity is:
\begin{equation}
	L_{c,j} \approx \frac{2 E_{c,j} c}{R_* \theta_{c,j}^2} = 4 \times 10^{51} ~E_{51.5}  R_{11}^{-1}  \theta_{cj,0.5}^{-2}  \ ,
	~{\rm \frac{erg}{s}} ~~~;~~~{\rm no~mixing}
\end{equation}
where $\theta_{cj,x}=\theta_{c,j}/(x \rm{~rad})$. The observed temperature is 
\begin{equation}
	T_{c,j} \sim 100 {\rm~ keV}~~~;~~~{\rm no~mixing.}
\end{equation}

Thus, if there is no mixing every long GRB emits a relatively smooth, very bright, $\sim 100$ keV quasi-thermal pulse with a duration of several seconds over a relatively wide angle. This  very bright signal  falls right in the middle of the energy window of all GRB detectors, including BATSE, {\it Swift} and the GBM, and it should be detected out to high redshift. Moreover, this signal is emitted over an angle that is much larger than the GRB jet, and therefore GRBs with relativistic cocoon outflows that point towards Earth are more numerous by a factor $(\theta_{c,j}/\theta_{j})^2$ than  GRB jets that point towards Earth. Therefore if there was no mixing in the cocoon  many such pulses should have been detected, possibly even as many as observed GRBs. In reality GRB detectors do not detect many events with such characteristics, if any. We conclude that these observation (or rather lack of) rule out the scenario in which GRB jets produce a significant cocoon whose  shocked jet material does not undergo a  significant mixing. \\

\noindent {\it ii. Partial mixing ($1 \ll \eta_{c,j}<\eta_b$)}\\
In section \ref{sec:jetDynamics} we use the fireball solution to approximate the evolution if the shocked jet material has $\eta_{c,j} \gg 1$. We assume there that $\eta_{c,j}$ is homogenous. However, in the case of partial mixing, $\eta_{c,j}$ may vary within the shocked jet material, namely different fluid elements in the shocked jet have different values of $\eta$. Yet, we use the fireball solution to approximate the evolution of material with a given value of $\eta$, as if it is not affected by material with different values of $\eta$. We can do that since material that is on top, closer to the jet's head at the time of breakout, has less time to mix and therefore it is expected to have higher $\eta$ than material near the jet base. As shown in figure \ref{fig:eta}, simulations support this expectation. Hence, the lighter and faster material on the top is free to expand first while the heavier material at the bottom which expands more slowly is also free to expand later, after the fast material has already evacuated the cocoon cavity. Using the fireball approximation all we need  to calculate the emission from material with a given value of $\eta$ is the fraction of the cocoon's energy (and volume) that it carries (occupies). {Since the energy density before the breakout is roughly uniform in the cocoon, the energy and volume fractions are similar}. We parametrize this fraction by $f_\Gamma$  so that $E(\eta) 
d \log \eta \equiv f_\Gamma E_c$. 
At this regime, under the fireball approximation, the final Lorentz factor of each fluid element is $\Gamma=\eta$ and hence this is also the fraction of energy in a given logarithmic interval of $\Gamma$. If, as suggested by simulations, the energy is divided uniformly for every logarithmic scale of $\Gamma\beta$ in the range $0.1-10$ then $f_\Gamma \sim 0.1 $ for every $\Gamma$ in this range. We therefore use the notation $f_{\Gamma,0.1}=f_\Gamma/0.1$ as the canonical value of $f_\Gamma$.

In case that $1 \ll \eta_{c,j} < \eta_b$ the  radiation remains trapped also during the coasting phase (after the outflow stops being accelerated) and its luminosity drops significantly. In this case the  luminosity depends also on the evolution during the coasting phase. At the beginning of the this phase the outflow maintains a constant width but later it starts spreading.  This change in the evolution dictates another critical value of the baryonic loading\footnote{The critical baryonic loading denoted here $\eta_s$ is denoted in  \cite{Nakar05} as $\eta_4$.} \citep{Nakar05}:
\begin{equation}
	\eta_s = 50 ~E_{51.5}^{1/5} \theta_{10^o}^{-2/5} R_{11}^{-2/5} f_{\Gamma,0.1}^{-1/5},
\end{equation} 
where we approximate the width of the relativistic shell before it starts expanding as a fraction $f_{\Gamma}$ of the size $R_*$, namely as $f_{\Gamma}R_*$.
For $\eta_s < \eta_{c,j} < \eta_b$ the radiation decouples during the costing phase, but before spreading starts while for  $\eta_{c,j} < \eta_s$ decoupling occurs only after spreading starts. 

Here we give only the solution for the case $\eta_{c,j} < \eta_s$, since this seems to be the more relevant one 
for the problem at hand. Also, since $\eta_s \lesssim \eta_{BB}$ for typical GRB parameters we assume that the 
radiation is in thermal equilibrium at the time of the breakout. Namely, the initial temperature of the outflow 
is $T_{BB,0}$. The radiation is released at the photospheric radius \citep{Nakar05}:
\begin{equation}
	\begin{array}{lll}
	R_{ph}&\approx& \left( \frac{\kappa E_c f_\Gamma}{2 \pi \theta_{c,j}^2 c^2 \Gamma} \right)^{1/2}\\ 
	&&\\
	&\approx& 7 \times 10^{13} ~E_{51.5}^{1/2} 	\theta_{cj,0.5}^{-1} f_{\Gamma,0.1}^{1/2} \Gamma_{10}^{-1/2}  ~{\rm cm} \ ,
	\end{array}
\end{equation} 
where $\Gamma_x=\Gamma/x$ and we took $\kappa=0.2 {\rm~gr/cm^2}$. The observed duration is therefore:
\begin{equation}\label{eq:tcjR}
	\begin{array}{lll}
	t_{cj,R} &\approx& \frac{R_{ph}}{2c\Gamma^2} \\ 
	&&\\
			&\approx& 10 ~E_{51.5}^{1/2} \theta_{cj,0.5}^{-1} f_{\Gamma,0.1}^{1/2} \Gamma_{10}^{-5/2}  ~{\rm s} \ .
	\end{array}
\end{equation}
The volume of the shell at the photosphere in the comoving frame is $V'_{ph} \sim 2\pi \theta_{cj}^2  R_{ph}^3/\Gamma$.  
This should be compared to $V_c f_\Gamma$, the initial volume occupied by the material with baryonic loading $\Gamma$.  
The temperature and energy in the observer frame is reduced by a factor of $(V_c f_\Gamma/V'_{ph})^{1/3}\Gamma$. Thus, the 
 {\it isotropic equivalent} luminosity and the observed temperature are: 
\begin{equation} \label{eq:LcjR}
	\begin{array}{lll}
	L_{cj,R} &\approx& \frac{2 E_c f_\Gamma }{t_{obs} \theta_{cj}^2}\left(\frac{V_c f_\Gamma}{V'_{ph}} \right)^{1/3} \Gamma \\ 
	&&\\
	&\approx& 1.3 \times 10^{48}  \theta_{10^o}^{2/3} R_{11} \theta_{cj,0.5}^{-2/3} f_{\Gamma,0.1}^{1/3} \Gamma_{10}^{13/3}  ~{\rm erg/s,} 
	\end{array}
\end{equation}
and 
\begin{equation}\label{eq:TcjR}
	\begin{array}{lll}
	T_{cj,R} &\approx& T_{BB,0} \left(\frac{V_c f_\Gamma}{V'_{ph}} \right)^{1/3} \Gamma \\
	&&\\
	 &\approx& 130 ~E_{51.5}^{-1/4} \theta_{10^o}^{1/6} R_{11}^{1/4} \theta_{cj,0.5}^{1/3} f_{\Gamma,0.1}^{-1/6} \Gamma_{10}^{11/6}  ~{\rm eV.} 
	\end{array}
\end{equation}
If the energy per logarithmic scale of $\Gamma$ is constant (namely, $f_\Gamma$ is constant) than $L_{cj,R} 
\propto t^{-26/15}$ and  $T_{cj,R} \propto t^{-11/15}$. This implies that the contribution of the cooling 
emission from partially mixed relativistic jet material is bright in X-rays only a for a very short time  
($\lesssim 100$ s). However, it can be significant in the UV and to some extent in the optical. The 
temperatures during the relativistic phase are $\gg 10,000$ K and therefor these bands are in the 
Rayleigh-Jeans part of the spectrum and the luminosity in a given UV/optical band increase with time 
as $L_{UV/opt} \propto L/T^3 \propto t^{7/15}$. 

\begin{figure}[!t]
\epsscale{1.23} 
\includegraphics[width=60mm,angle=90.]{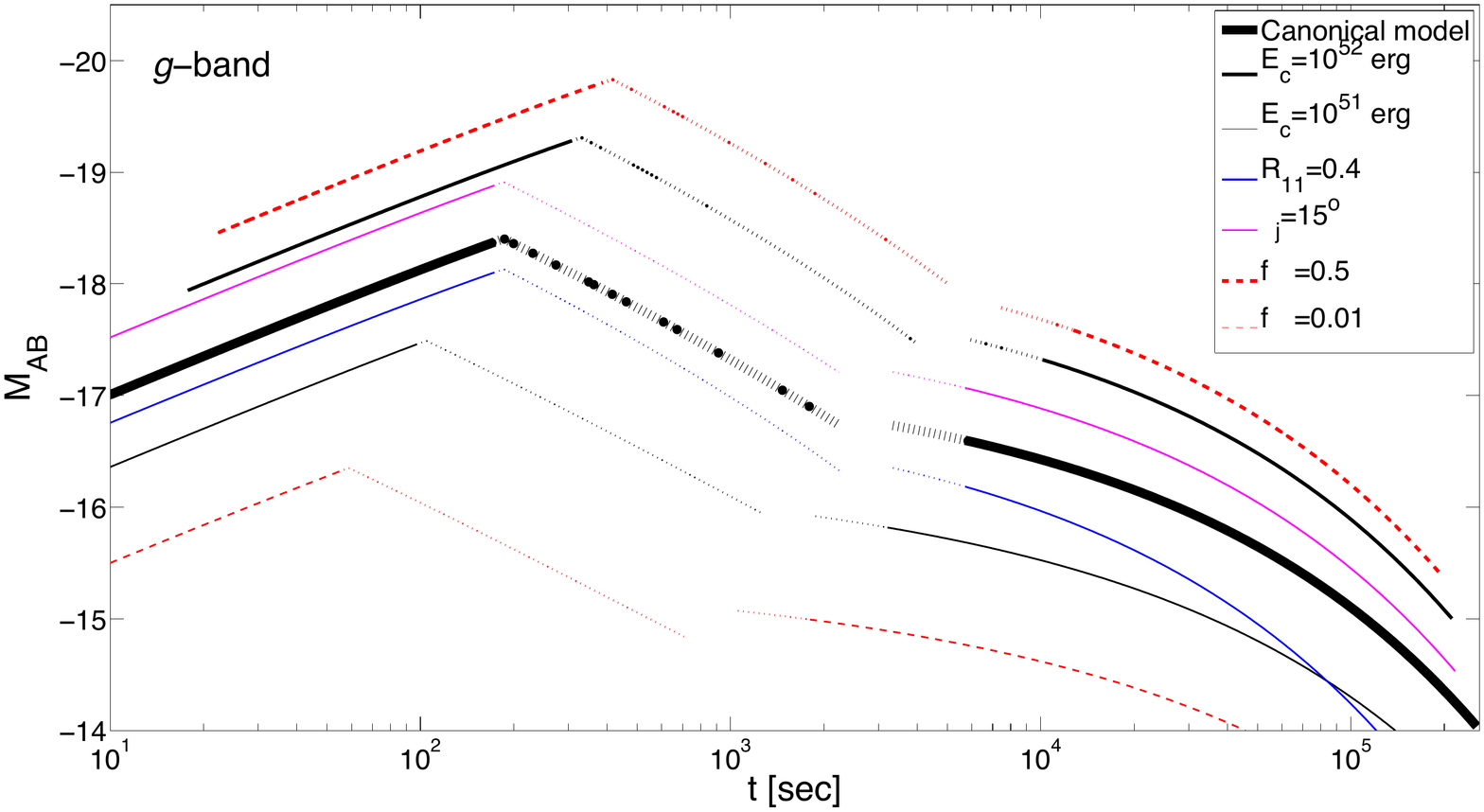}
\caption{{The optical light curve for different parameters, assuming a partial mixing that distributes a constant amount of energy per logarithmic scale of $\Gamma \beta$. The rising part at early time ($t < 10^3$ s) is the contribution from the relativistic component (solid lines). It peaks at $\Gamma=3$. This emission is beamed to an opening angle $\theta_{c,j} \approx 0.5$ rad. The late time decaying emission arises from the Newtonian ($\Gamma \beta <1$) component (solid lines). This contribution is isotropic. The contribution of the mildly relativistic component ($1 \lesssim \Gamma \beta \lesssim 3$), that is more difficult to calculate precisely is marked with dotted lines.  
The canonical model ($E_{51.5}=\theta_{10^o}= R_{11}=\theta_{cj,0.5}=1$ and $f_{\Gamma}=f_{\beta \Gamma,1}=0.1$) is shown with a thick line. Other lines show various deviations from this model. Note that the two cases of $f_{\Gamma\beta}=0.5$ and $f_{\Gamma\beta}=0.01$ (dashed lines) are not physical. The purpose of these curves is to illustrate by how much the light curve vary when it is dominated by material with $\Gamma\beta$ that carries more or less than 0.1 of the total cocoon energy (see text). }}
\label{fig:CoolingOpt}
\end{figure}

\begin{figure}[!t]
\epsscale{1.2}
\includegraphics[width=60mm,angle=90.]{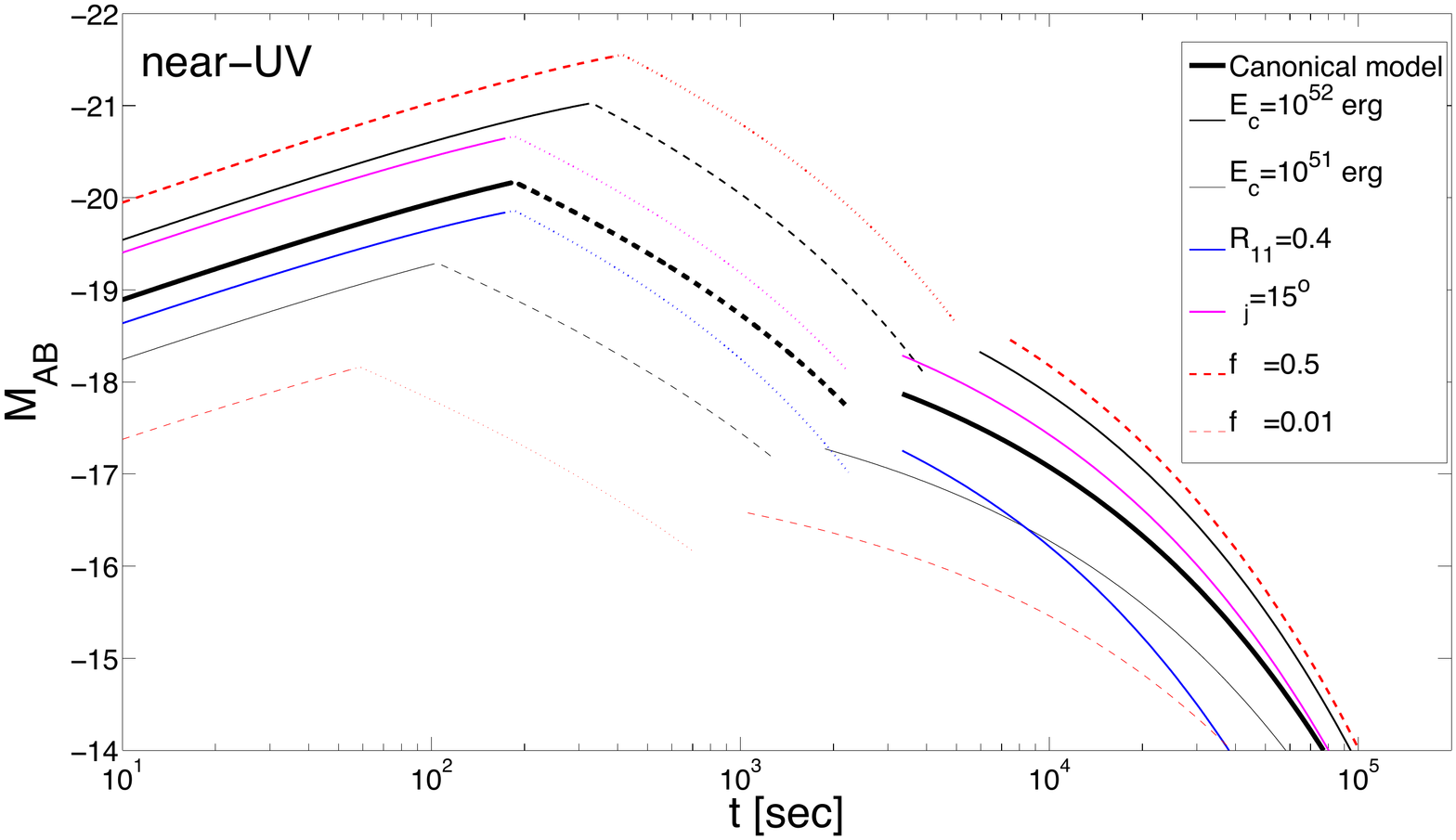}
\caption{Same as Fig. \ref{fig:CoolingOpt} for the NUV emission. }  
\label{fig:CoolingUV}
\end{figure}

Examples of the predicted cooling emission light curves from the shocked jet material in the optical and UV are shown in figures \ref{fig:CoolingOpt} and \ref{fig:CoolingUV}. The figures depicts both the relativistic and the Newtonian regimes (solid lines). The fully relativistic regime is calculated using equations \ref{eq:tcjR}-\ref{eq:TcjR} by varying $\Gamma$ from 10 to 3 and keeping a constant $f_{\Gamma}$. It generates the rising part of the light curves at early time and the peak is observed when the emission is dominated by material with $\Gamma \approx 3$. This emission is beamed into a wide beam with an opening angle $\theta_{c,j}$ taken here as $0.5$ rad. The Newtonian regime is plotted by taking the values from equations \ref{eq:tcjN}-\ref{eq:TcjN} and evolving $L$ and $T$ according to equations \ref{eq:LcjN_evolution} and \ref{eq:TcjN_evolution} with $s=1$ (equal amount of energy per logarithmic velocity scale). The emission in this regime is isotropic.

None of the two regimes describes accurately the emission from the mildly relativistic material. Yet, we plot an approximation of this emission in dotted lines. To do that we extend the Newtonian phase up to $\beta=1$ and the relativistic regime down to $\Gamma=1$. In the extension of the relativistic case we also use equations \ref{eq:tcjR}-\ref{eq:TcjR}, but with a smooth transition of $\theta_{c,j}$ and $t_{cj,R}$ from their relativistic limit (0.5 rad and $R_{ph}/2c\Gamma^2$, respectively) at $\Gamma=3$ to their Newtonian limits ($\pi/2$ rad and $R_{ph}/c\Gamma^2$) at $\Gamma=1$. The relativistic  extension to $\Gamma=1$ and the Newtonian extension to $\beta=1$ do not connect, although the physical light curve is continuous, this is due to the inaccuracy of both approximations at the mildly relativistic regime.    

Figures \ref{fig:CoolingOpt} and \ref{fig:CoolingUV} show also the effects of some of the parameters on the optical and UV light curves. It includes what we consider as the canonical
model ($E_{51.5}=\theta_{10^o}= R_{11}=\theta_{cj,0.5}=1$ and $f_{\Gamma}=f_{\beta \Gamma,1}=0.1$) as well as models where we vary one of the parameters at a time (e.g., $R_*$, $\theta_j$). This provides an idea of by how 
much the signal can vary from one event to another. We also include two cases where the fraction of energy per logarithmic scale (denoted in the figure for short as $f_{\beta\Gamma}$) is taken to be $f_{\Gamma}=f_{\beta \Gamma,1}=0.5$ and $f_{\Gamma}=f_{\beta \Gamma,1}=0.01$. {Note that these two cases are not physical, as the first contains too much energy and the second contains too little. Namely, if  $f_{\beta\Gamma}=0.5$ [$f_{\beta\Gamma}=0.01$] at some value of $\beta\Gamma$ it cannot be constant. It must be lower [higher] at some other value of $\beta\Gamma$. The purpose of these curves is to illustrate by how much the light curve varies when it is dominated by material with $\beta\Gamma$ that carries more or less than 0.1 of the total cocoon energy. }

\subsection{The Cocoon's afterglow}
The interaction of the cocoon's outflow  with the external medium is another source of emission. A significant contribution is expected only from interaction of the relativistic component, if such one exists. We  consider, therefore,  only the contribution of the shocked jet material in the case that at least part of it expands relativistically. The beam of the shocked jet material is much broader than the  narrowly collimated and highly relativistic GRB jet. {Therefore, at viewing angles that are much larger than the jet opening angle, the cocoon and not the GRB jet (i.e. the standard orphan afterglow) dominates the observed emission at least during the first several days}. This happens  until the Lorentz factor  of the GRB afterglow  drops to one over the viewing angle, at which stage the orphan afterglow signal dominates. Therefore, at angles larger than the jet's opening angle, the GRB jet can be ignored at early times and the theory of the cocoon afterglow emission is similar to that of regular GRB afterglow \citep[][ and references therein]{Piran04}. 

If a slower moving material carries more energy than faster moving material, the calculation must include a continuous energy injection. If instead the fast moving material carries a significant fraction of the outflow energy (as in the case of a constant energy per logarithmic scale of $\Gamma \beta$), then the emission will be dominated by the interaction of the fastest moving material and energy injection can be ignored. Here we consider such a case and estimate the cocoon afterglow emission by considering only the interaction of the fastest moving material that carries a significant fraction of the cocoon's energy. As in previous sections $\Gamma$ is  the characteristic Lorentz factor of this material,  $f_\Gamma$ is the fraction of the total cocoon energy that it carries   and $\theta_{c,j}$  is its half-opening angle. 

For a given values of $E_c$, $\Gamma$,  $f_\Gamma$, $\theta_{c,j}$, external density distribution and the usual microphysics parametrization one can use the standard afterglow theory to calculate the predicted emission. Here we will use a different approach and estimate this emission  by scaling actual observations of regular GRB afterglows to the conditions expected here. Since the cocoon and the jet propagate into the same external medium we expect the external density distribution and microphysics parameters to be the same. Therefore, the only differences between the regular GRB   afterglow (generated by the jet)  and one generated by the shocked jet cocoon arises due to the differences in the isotropic equivalent energies and in the initial Lorentz factors. 

The peak of the cocoon afterglow emission is observed at $t_{c,aft}$,  once the cocoon's material reaches the deceleration radius and begins to slow down, This happens at 
\begin{equation}
	t_{c,aft} \approx 0.3 ~ \left ( \frac{2 E_{51.5} f_{\Gamma,0.1}}{n \theta_{cj,0.5}^{2}} \right )^{1/3} \Gamma_{10}^{-8/3} {\rm ~d} \ , 
\end{equation}  
 for a constant external density, $n$;  and at 
\begin{equation}
	t_{c,aft} \approx 0.01 ~ \left ( \frac{2 E_{51.5} f_{\Gamma,0.1}}{A_* \theta_{cj,0.5}^{2}} \right ) \Gamma_{10}^{-4} {\rm ~d} \ , 
\end{equation}
for  a wind density profile $\rho \propto A r^{-2}$  
where $A_* \equiv A/(5 \cdot 10^{-11} {\rm~gr/cm})$.

We estimate the luminosity  at a given time after the peak 
 by  comparing it  with the luminosity of observed GRB afterglows at the same time. The ratio of the isotropic equivalent energies of   the fastest moving cocoon material and the jet is $\sim f_\Gamma (\theta_j/\theta_{c,j})^2$. The optical, UV and X-ray luminosities of a GRB afterglow at a given time are roughly linear in the isotropic equivalent energy of the outflow both for a constant density and a wind  \citep[e.g.][]{GranotSari02}). Therefore, at $t>t_{c,aft}$ and for a viewing angle larger than $\theta_j$ but smaller than $\theta_{c,j}$, the cocoon afterglow luminosity in these bands can be estimated as:
\begin{equation}
\label{eq:Lcaft}
	L_{c,aft} \sim 0.01 L_{j,aft} \left( \frac{\theta_{10^o}}{\theta_{cj,0.5}}\right)^2  f_{\Gamma,0.1}
\end{equation}  
where $L_{j,aft}$ is the regular on-axis GRB afterglow observed by an observer with a viewing angle within the opening angle of the jet. For our canonical parameters, the cocoon afterglows peaks after a fraction of a day and it is about 100 times fainter than a regular GRB afterglow. However, its radiation is emitted over a solid angle that is larger by a factor of $\sim 10$ than the jet's solid angle. {To estimate  the detectability  of cocoon afterglows in soft X-rays we use the observed GRB afterglows after 1 day that typically have a luminosity of $\sim 10^{46}$ erg/s \citep{Margutti13}. This implies that for our canonical parameters the X-ray luminosity of a typical cocoon afterglow at that time is $\sim 10^{44}$ erg/s. To estimate  the luminosity of optical cocoon afterglows we compare it to observed GRB afterglows after 1 day that  typically have an absolute optical magnitude in the range  $-21$ to $-25$ \citep{Kann11}. Therefore the optical emission from cocoon afterglows after 1 day is expected to be in the range $-16$ to $-20$. Below, when estimating detectability, we use a value of $-18$ as the canonical absolute magnitude of cocoon afterglow at 1 day.}

\section{Detectability}
\label{sec:detectability} 
We turn now to discuss the detectability of the resulting signals by some of the present and future detectors. 
For brevity we discuss the detectability only for the canonical model. 
We note that when the detectable signal is generated by the shocked stellar material our predication is more robust, and when it is generated by the shocked jet material our prediction depends on the mixing that is not well constrained. As discussed above, our canonical model for the mixing assumes that the shocked jet cocoon energy is distributed uniformly for each logarithmic scale of $\Gamma\beta$. 
The reader can easily scale the result to other possible values using Figs. \ref{fig:CoolingOpt} and \ref{fig:CoolingUV} and Eqs. 
\ref{eq:tcs}-\ref{eq:Tcs}, \ref{eq:tcjN}-\ref{eq:TcjN} and  \ref{eq:tcjR}-\ref{eq:TcjR}.  A change by one magnitude of either the source's strength or the detector's sensitivity will change the number of detected events by a factor of $\sim 4$. 
We consider detectors that are operating at 100\% of the time and we  neglect possible obscuration or absorption of the signals. These effects could reduce the idealized observed event rates discussed below. 

We ignore in the observed rates estimated here the cases of  cocoon emission from choked jets. 
We expect the characteristics of the mixing to be different,  probably much
more effective, as the relativistic cocoon has still to cross the rest of the stellar envelope before emerging. However, the rate of these events will probably be much larger. 
These events  would  almost certainly produce an observable Newtonian signatures and possibly more. These could significantly increase the observed rates. 

A  detector with a limiting magnitude $m_{det}$ can detect a transient source with an absolute magnitude $M$   up to a distance: $ D_{max} = 10^{(m_{det } - M-40)/{5}}$ Gpc .
For simplicity we have neglected cosmological redshift effects as most of the events discussed here can be detected for distances of order $\sim 1$ Gpc. 
We take  $R_{GRBs} \approx 1 {\rm Gpc}^{-3} {\rm yr}^{-1} $ as the local rate of observed LGRBs, i.e., GRBs with jet that points towards us \citep[e.g.,][]{Wanderman10}. We  assume a beaming factor of   $f_{B_{GRB}} =70$, 
corresponding to a typical LGRB jet  opening angle of $\sim 10^o$. 
If the emission lasts for a duration $t$ above the limiting magnitude and it is beamed into a cone with a half opening angle $\theta$, then the number of detectable events at any given moment per steradian is
\begin{equation}
N_1 =  \frac{1-cos(\theta)}{3} D_{max}^3 f_{B_{GRB}} R_{GRB} t  \ . 
\label{eq:N1} 
\end{equation} 
If the luminosity at the observed band  decreases like $t^{-\alpha}$, the number of event behaves as $t^{1-3\alpha/2}$. Unless the light curve is flatter than $-2/3$ the detectability is dominated by the peak luminosity. Otherwise it is dominated by the latter longer signal.
If the survey cadence is longer than $t$, then the detection rate is simply $N_1$ multiplied by the accumulated area (including multiple visits) per unit of time covered by the survey. If instead the survey cadence  is shorter than $t$  then all the events in the covered area, $S$, during the time of the survey are detected and the detection rate is $S N_1/t$.

We turn now to consider the detectability of the specific components discussed earlier.
We summarize in table \ref{tab:signatures}  the different contributions that we consider. 

\begin{table}[t]
\caption[]{Different signatures of the various components.
\\ }
\label{tab:signatures}
\scalebox{0.9}{\begin{tabular}{lccccc}
\hline \hline
Component & Band                    & Beaming & $\Gamma\beta^\dagger$ &Luminosity$^\ddagger$   & Duration \\ \hline
{\bf Shocked stellar} & &&& &     \\ \hline 
				&opt       	& isotropic & 0.1 	 & -16   			& a day \\ \hline \hline
{\bf Shocked jet}&&&&& \\ \hline                                    
Full mixing 	& opt       & isotropic & 0.1 	& -16       		& a day       \\ \hline 
No mixing$^*$ 		& 100 keV   &  0.5 rad  & $>$100&$4 \times 10^{51}$	& 3 s           \\ \hline
Partial mixing: &&&&& \\                                            
Newtonian cooling & UV  	& isotropic & 1		&  -18         		& an hour           \\
   		 		& opt  		& isotropic & 1 	&  -17         		& an hour           \\ 
Rel. cooling 	& UV		&  0.5 rad  & 3 	& -20    			& 200 s  \\
		 		& opt       & 0.5 rad   & 3 	& -18     			& 200 s           \\ 
Rel. afterglow 	& X-ray     & 0.5 rad   & 10 	& $10^{44}$			& a day        \\
 				& opt/UV    & 0.5 rad   & 10 	& -18     			& a day        \\
\hline \hline\\
\end{tabular}}
{\scriptsize \\
$\dagger$ The $\Gamma\beta$ of the material that dominates the luminosity of that component at the peak. 
$\ddagger$ Peak luminosity; Optical/UV in absolute AB magnitude, $\gamma$-rays and X-rays in erg/s.\\
$^*$ Ruled out by observations
\\	
{The peak luminosity at various bands emitted by different cocoon components for our canonical model. The shocked stellar material signal is the most robust one. The shocked jet material signal depends on the unknown mixing. We consider here three options, full, partial and no  mixing. With full mixing there is no difference between the shocked stellar and shocked jet material. No mixing is ruled out by observations. We consider partial mixing to be the most likely case, as it is supported by simulations. Here we consider partial mixing that distributes the shocked jet energy uniformly for every logarithmic scale of $\Gamma\beta$ in the range $0.1-10$. Thus any of the components that contributes to the partial mixing case carries about 10\% of the total cocoon energy.  }
} 

\end{table}

\subsection{ $\gamma$-rays} 
A soft $\gamma$-ray/hard X-ray signal is expected from a  relativistic unmixed jet cocoon. It has $\sim100$ keV quasi-thermal spectrum, a duration of a few seconds and an extremely bright luminosity over a relatively wide opening angle of half a radian.   With a typical luminosity of  $\gtrsim 10^{51}$ ergs/s such a signal could be easily detectable by {\it Swift}, for example,  up to $z \gtrsim 1$. Given that the beaming angle is significantly larger than the beaming angle of a typical GRB jet, we would have detected a large number  of  such events with a relatively soft quasi-thermal emission. 
As  such signals have rarely, if ever, been  detected the possibility of no mixing is ruled out observationally.
Note that these signals are much shorter and much brighter than {\it ll}GRBs.

\subsection{ X-rays} 
A detectable X-ray signal is  expected only if a significant fraction of the shocked jet cocoon energy is deposited in material with a Lorentz factor $\gtrsim 10$. In this case  the cocoon X-ray afterglow is beamed to a half opening angle of about $0.5$ rad and its luminosity, at a day after the burst,   is $\sim 10^{44}$ erg/s. ISS-Lobster is a proposed wide field-of-view soft X-ray (0.3-5 keV) detector that will scan a large fraction of the sky ($\sim 20\%$) every ISS orbit down to a limiting flux of $\approx 10^{-11} {\rm~ erg/s/cm^2}$ \citep{Camp13}. Our predicted canonical X-ray cocoon afterglow would be detected  by ISS-Lobster out to a distance of $\approx 300$ Mpc, implying a detection rate of about one such event every year.

\subsection{ UV} 
The strongest UV signal arises from the relativistic component that has a typical Lorentz factor of $\sim 3$. This signal has a duration $\sim 200$ sec, an absolute near-UV magnitude of $\sim -20$ and an opening angle of about $0.5$ rad. The mildly relativistic ($\Gamma\beta \approx 1$) component produces a signal that is slightly fainter (-18), but it lasts  longer ($\sim$ 1hr) and it is roughly isotropic. 
ULTRASAT \citep{Sagiv14} is a proposed 
near-UV transient mission that will continuously observe a region of 200 sq degrees with a limiting magnitude of 21.5 and a data collection time of 900 s (appropriate for the mildly relativistic isotropic signal) . Data collection can be taken also at 300 s and then the limiting magnitude is 21 (appropriate for the relativistic signal). {Using the above estimates we find that ULTRASAT will detect the relativistic signal of a canonical cocoon cooling emission out to $\sim 1.5$ Gpc  and the mildly relativistic  signal out to $\sim 800$ Mpc. Given that the brighter signal is beamed, while the fainter one is isotropic, the detection rates of both signals are comparable, about one event per year. }

\subsection{ Optical} 

\subsubsection{ Shocked stellar matter} 
There are several signals in the optical band. First we consider the contribution from the shocked stellar material as it is the most robust among the different optical signals. This is an isotropic component with a luminosity of $\sim 10^{42}$erg/s corresponding to an absolute optical magnitude of -16 and a duration of about a day.  We consider the detectability of these events using two telescopes, ZTF and LSST. The ZTF has a limiting magnitude of 20.5 and it covers about a quarter of the sky  with a cadence of once per night using a tiling of $50 {\rm~ deg^2}$  and $\sim 1$ min per pointing \citep{Bellm14}. ZTF will detect the shocked stellar emission out to a distance of about $200$Mpc at a rate of about one event per year. LSST that will become operational in the early 20ies, will cover the whole observable sky down to 24'th magnitude  once every 3 days \citep{LSST09}.  The LSST will detect the shocked stellar emission out to about a Gpc, detecting  about one such event per week.

\subsubsection{Relativistic jet afterglow} 
The strongest optical signal arises from the afterglow like component driven by the relativistic (but mixed) jet material. This signal peaks at a magnitude of about -18 with a duration of about a day. This signal can be detected out to a distance that is larger by a factor of 2.5 compared to the signal of the shocked stelar cocoon, but it is beamed to within an opening angle of about $0.5$ rad compared to the isotropic shocked stellar signal. Therefore the detection rate of optical cocoon afterglows in our canonical model is twice the detection rate of the shocked stellar cocoon optical signal.

 \subsubsection{Relativistic Cooling Cocoon} The optical contribution of the Newtonian part (late time) of the relativistic outflow is weaker by about half a magnitude than the contribution from the shocked stellar material. Hence we can ignore this contribution. The contribution from the relativistic part is however more interesting. 
{It is brighter and it can reach -20th magnitude for the canonical case. However it is very short, lasting about 200 sec. Since optical surveys cover a very small fraction of the sky within this time the probability to detect this signal by ZTF is small. The LSST may detect this short signal about once per year.}

\subsection{Identification} We have considered here a single detection of a transient source. How can this source be identified and related to this phenomena? 
{The  long GRBs, that we discuss here,  are followed by powerful  broad-line type Ic SNe.  The optical signals of those SNe peak about two weeks after the explosion at an absolute magnitude of -18 to -19. The detection of such a unique SN, two weeks after the cocoon emission, will confirm its nature. Moreover,  due to their brightness and longer duration the SNe are easier to detect than the cocoons. Hence, once such an SN  has been identified one can go back and search the earlier data at the same point for the cocoon emission.
Finally, we note that at late time the orphan afterglow from the GRB jet becomes brighter than the cocoon emission. If the viewing angle is not too large (relative to the GRB's jet opening angle) the orphan afterglow signal  may  be detected as well. }  

\section{Cocoon emission from short GRBs}
\label{sec:SGRBs}
The formation of a cocoon is inevitable if the GRB jet is  launched inside a star, as in the Collapsar  model for LGRBs. But, an energetic cocoon may also form in short GRBs (SGRBs), if those are   generated following the coalescence of two neutron stars
\citep{Nagakura14,Murguia-Berthier14,MurguiaBerthier16}.  A large body of work shows that a significant amount of material ($M_{ej}\sim 0.01 M_\odot$) is ejected during the last stages of the in-spiral and the merger itself \citep[see e.g. table 1,  Fig. 1 and the subsequent discussion in][]{Hotokezaka15}. In particular in addition to the dynamical ejecta ejected during the last phases of the in-spiral, a significant wind is expected also in case that the merger leads to the formation of a hypermassive neutron star (HMNS) \citep{Perego14,Siegel14}. The formation of HMNS can also lead to a delay between the merger and the launching of the jets, if those are ejected only following the collapse of the HMNS to a black hole. The propagation of SGRB jets through the ejected material in this scenario was recently explored by \cite{Nagakura14} and \cite{MurguiaBerthier16}. These studies show that under reasonable assumptions the jet propagates through an effective atmosphere of $\sim 0.01 M_\odot$ at mildly relativistic velocities and it breaks out  at a radius of $\sim 10^{9} - 10^{10}$ cm. The duration of the jet propagation within the ejecta is comparable to a duration of the subsequent SGRB, implying that, just like in LGRBs,  the cocoon and the GRB energies are comparable. Therefore, we can apply our model of the cocoon emission also to SGRBs.   

The uncertainty concerning typical SGRB parameters is much larger than the uncertainty concerning LGRBs  \citep[see][for a review]{Nakar07,Berger14}).   Typical values of the isotropic equivalent energies are: $10^{50} -10^{52}$ erg. The jet opening angle is highly uncertain. We will use here a typical value of   $10^o $, with which the corresponding energies are $10^{48}  -  10^{50}$ erg. The opacity of the  cocoon's matter  is also not well constrained. Depending on the amount of r-process elements it contain and their maximal atomic number, $\kappa$ can be in the range $0.1-10 {\rm~cm^2/gr}$. Here, following \cite{Perego14} we use as canonical value for SGRB cocoons $\kappa=1 {\rm~cm^2/gr}$ which they find appropriate to material at high latitude, mostly wind material, in which the jet propagates.
Plugging the typical values  into equations \ref{eq:tcs} - \ref{eq:Tcs} we find that a detectable cooling emission is expected only in case of  favorable parameters. For example, taking  $E_c=10^{50}$ erg, and  a breakout radius of $10^{10}$ cm the cooling emission from the shocked surrounding material peaks after several hours at a luminosity of $\sim 10^{41}$ erg/s and a temperature of $\sim 10,000$ K. This corresponds to an absolute optical magnitude of about $-14$. 

The comparison of the cocoon afterglow to the GRB afterglow in SGRBs is similar to the comparison we carried out for LGRBs. Thus, assuming that the cocoon and the SGRB have similar energies, and taking the same canonical values we took for LGRBs in equation \ref{eq:Lcaft}, we expect the cocoon afterglow to be two orders of magnitude fainter than the SGRB afterglow. Typical SGRB afterglows have luminosities of $\sim 10^{43}$ erg/s in the optical after seven hours \citep{Berger14}. Therefore, in this scenario the cocoon afterglow luminosity at this time is about $10^{41}$ erg/s, corresponding again to an absolute magnitude of $-14$. This signal is beamed to a half opening angle of about 0.5 rad. Just like in LGRBs, unless the observing angle is very close to the jet opening angle this signal will be stronger than the corresponding orphan afterglow of such a burst. 

Another signal arises in SGRB.  The cocoon shocked external material will be radioactive. This signal is analogous to the macronovae/kilonovae signal and therefore we denote it as the cocoon's macronova.
The time, luminosity and temperature at the peak can be estimated using equations \ref{eq:mcs} and \ref{eq:vcs} for $m_{c,s}$ and $v_{c,s}$ and adding a radioactive energy injection rate per unit of mass $\dot{\epsilon}$. Similarly to the cooling emission the peak is observed when $\tau \approx v_{c,s}/c$ and thus the peak time, $t_{MN}$, can be estimated using equation \ref{eq:tcs} (replacing the progenitor mass by $M_{ej}$). The luminosity can be approximated as the instantaneous energy injection at $t_{MN}$, namely $m_{c,s} \dot{\epsilon}(t_{MN})$. This approximation yields:
\begin{equation}\label{eq:Lmn}
		L_{MN} \sim  4 \times 10^{40} ~E_{49}^{0.325} \theta_{10}^{0.05} M_{ej,-2}^{0.025} \kappa_1^{-0.65} \frac{\dot{\epsilon}}{\dot{\epsilon}_0} ~{\rm \frac{erg}{s}} ,
\end{equation}
where $E_{49}=E_c/10^{49}$ erg, $M_{ej,-2}=M_{ej}/10^{-2}M_\odot$ and $\dot{\epsilon}_0=10^{10}(t/day)^{-1.3} {\rm~erg/gr/s}$. The observed temperature is roughly
\begin{equation}\label{eq:Tmn}
		T_{MN}  \sim 11,000 ~E_{49}^{-0.04} \theta_{10}^{-0.24} M_{ej,-2}^{-0.12} \kappa_1^{-0.41} \left(\frac{\dot{\epsilon}}{\dot{\epsilon}_0}\right)^{1/4} ~{\rm K}
\end{equation}
Interestingly, only $t_{MN}$ depends strongly on $M_{ej}$ and $\theta_j$ while $L_{MN}$ is practically independent of them and only weakly dependent on $E_{c}$. $T_{MN}$ depends weakly on all these parameters.

To estimate the actual signal we follow the results of \cite{Perego14}. They find that at high latitude about $2 \times 10^{-3}M_\odot$ are ejected into an opening angle of $40^\circ$ prior to the HMNS collapse. This corresponds to an isotropic equivalent ejected mass of $0.0085 M_\odot$. Since the high latitude ejecta is composed of light r-process material they estimate its opacity as $\kappa=1 {\rm~cm^2/gr}$. Plugging these values into equation \ref{eq:tcs} and assuming $\theta_j=10^\circ$ and $E_c=10^{49}$ erg we find that the signal peaks around $0.15$ day. At that time \cite{Perego14} find that $\dot{\epsilon} \approx 2 \dot{\epsilon}_0$. Including this injection rate in equations \ref{eq:Lmn} and \ref{eq:Tmn} we obtain a luminosity of about $8 \times 10^{40}$ erg/s at a temperature of about $13,000$ K,  corresponding to an optical magnitude of about $-13$. A cocoon energy of $10^{50}$ erg generates a signal that is brighter by about 1 mag.  This cocoon macronova signal is brighter in the optical than the main high latitude macronova event calculated by \cite{Perego14} and it is isotropic.

These signals are weak, corresponding to an apparent  magnitude 22.5-23.5 for 
a source at a distance of  200 Mpc. However they are  comparable or even stronger than typical IR macronovae/kilonovae signals that are the hallmark of search for optical EM counterparts for gravitational wave events from mergers \citep{Barens13,Tanaka13},  \citep[see][for recent reviews]{Fernandez15,Tanaka16}. In particular these signals are also much bluer. 
Therefore the search should be in the optical which is much easier than a search in the IR. However, given the short duration the search must be much faster. 
 
 \section{Conclusions}
\label{sec:conclusions}

As a long GRB jet propagates inside a stellar envelope its energy is deposited in the formation of a cocoon that surrounds it until it breaks out. The cocoon is made out of two parts - an inner region made of shocked jet matter and an outer region composed of shocked stellar material.  Both contain a comparable amounts of energy, but their masses are most likely very different. After breakout the cocoon expands freely and accelerates. The heavy shocked stellar material expands spherically and accelerates to Newtonian velocities. The shocked jet material terminal velocity depends on its mass, which in turn depend on the mixing level between jet and stellar material in the cocoon. If there is no mixing the shocked jet material accelerates to extreme relativistic velocities, while if there is full mixing the shocked jet and shocked stellar material expands together to Newtonian velocities.  Numerical simulations suggest that the situation is probably somewhere in between and that the jet material is mixed to different levels, where material near the jet base is more mixed than material near the head of the jet. The level of mixing is currently unknown and it is most likely depends on the exact properties of the jet and the progenitor. Preliminary results of simulations of hydrodynamic jets suggest that in these jets the energy is distributed roughly evenly per logarithmic scale of $\Gamma \beta$ (where $\Gamma$ is the terminal Lorentz factor after expansion) ranging from $\beta \approx 0.1$ to $\Gamma \approx 10$.

This cocoon structure suggests  three different observed cocoon signatures:
an isotropic emission from the cooling shocked stellar material, a beamed (over a wide angle) emission from the cooling relativistic shocked jet material and an afterglow like signature arising from the  interaction of the relativistic cocoon component and the surrounding matter. 
The first component, the cooling shocked stellar material, is practically inevitable. The Newtonian shocked matter expands spherically and emits its intrinsic energy once the radiation diffusion time becomes comparable to the dynamical time.  This isotropic optical signal has a duration scale of a days and it has a maximal absolute magnitude of  -16 for canonical parameters.  The signal depends on the cocoon properties at the time of breakout, which in turn depend on the properties of the jet and the progenitor. Thus, a detection can teach us about the progenitor properties as well as about the jet. In addition, being isotropic, this signal also probes the total rate of GRBs, and could in tern reveal what is their true beaming factor. This signal  might be enhanced by an additional possible contribution from the Newtonian jet if mixing is strong. 

The relativistic components of the shocked jet move faster and hence they  lead to an early, brighter  and bluer signature, which is beamed to a wide angle of around 0.5 rad. If there is no mixing the relativistic cocoon produces a short (a few seconds) extremely bright ($\gtrsim 10^{51}$ erg/s)  soft gamma-ray ($\sim 100 $ keV) quasi-thermal bursts. The lack of detection of such bursts in large numbers by GRB detectors such as BATSE, {\it Swift} and the GBM, rules out no mixing. This is an interesting result that may constrain the nature of the jet (e.g., matter or magnetic dominated). It is consistent with numerical simulations of hydrodynamic jets and it is yet be seen if it is also consistent with magnetized jets.

Partial mixing that deposits a significant fraction of the cocoon energy in material with Lorentz factor $\sim 2-10$, such as the one seen in numerical simulations, leads to a signal that peaks after less than a minute in the soft X-rays. Its luminosity and temperature drop with time, resulting in an increase in the UV/optical signal. The peak in these bands is seen after several hundred seconds, when material with  $\Gamma \approx 3$ dominates the emission. For our canonical parameters the UV [optical] peak has an absolute magnitude of -20 [-18]. 
 
The final contribution is the cocoon's  afterglow that arises from the interaction of the relativistic cocoon outflow with the surrounding material. It is dominated by the fastest material in the outflow that carries a significant amount of the cocoon energy. For $\Gamma \sim 10$  
the peak time of this emission is about an hour to a day and it is about two order of magnitude fainter than a regular GRB afterglow at the same time. It is beamed to a wide angle of around 0.5 rad and it is, therefore, $\sim$10 times more frequent than regular afterglows. If it exists, it is the the brightest and most detectable of all the cocoon's signals both in the optical and in X-rays (but not the UV). 

We have estimated the detection rates of cocoon emission by several future surveys. We considered only the canonical model and hence we provides only  rough estimates. The shocked stellar emission is the most promising optical signal (it is both robust and bright). It is predicted to be detected by the ZTF once per year and by the LSST once per week. The less robust (mixing dependent) optical afterglow emission is predicted to  be detected at slightly higher rates. The afterglow also predicts to produce a detectable X-ray emission. The proposed ISS-Lobster is predicted to detect one cocoon afterglow every year. Finally, a detectable UV signal is expected only from cooling emission of the relativistic  and mildly relativistic components. The proposed ULTRASAT telescope is predicted to detect a single event every year.

The cocoon signal is rather short (a day time scale) and therefore it will be challenging to identify it by itself. However, it is expected to be followed by additional components of the GRB/SN. First, if the viewing angle is not too large the so called orphan afterglow emission from the GRB jet should become dominant after several days or weeks. Second, the type Ic SNe that accompany LGRBs should be detectable a week or two after the explosion. Moreover, since these SNe are easier to detect than the cocoon emission, once such a SN  has been identified one can go back and search the earlier data at the same point for the cocoon emission.

Cocoon emission is also expected in failed GRBs, where the engine stops while the jets still propagates within the stellar envelope, and the jets are choked before breaking out of the envelope. We leave a detailed discussion of this case to a future work, however we can give here an educated guess for the expected signals from choked GRBs. If the jet is chocked after crossing a significant fraction of the envelope (about half way to the edge) the cocoon will break out successfully. The shocked stellar component in chocked jets is expected to be similar to that of successful jets. The choked jet material, which in this case spends a longer time in the cocoon before breakout, is expected to have an enhanced mixing. Thus we expect  the cocoons of chocked GRBs to have  Newtonian outflows similar to those of cocoons of successful GRBs,  but less relativistic outflows if any. In the future, if and when many cocoon signals will be detected the existence or absence of the signature from relativistic material may enable us to discriminate between choked and successful GRBs. Choked GRBs may be of special interest since their rate is most likely larger than that of successful GRBs \citep{Bromberg12}, and therefore they may significantly increase the detection rate of the optical cocoon signal.

Our discussion focused on cocoons that arise in LGRBs, where a cocoon is an inherent part of the Collapsar model.
However, matter ejected in a binary neutron star merger (either via a tidal interaction or due to neutrino or magnetic driven winds from a hypermassive neutron star) forms an envelope that a SGRB jet must penetrate. The interaction of the SGRB jet with this matter will produce a cocoon \citep{Nagakura14,MurguiaBerthier16}.  The main difference between this cocoon and the one arising in LGRBs is that its typical energy and the breakout radius are smaller by about two orders of magnitude, and hence the signal is much fainter.  Additionally,  in a neutron star merger the shocked envelope part of the cocoon will be
radioactive and the additional heating would increase its emission. 
The framework developed for the LGRB's cocoon can be applied to the signature of this one as well. We find  that the strongest signature of this cocoon will most likely be the afterglow signature from the relativistic part of the cocoon, if it exist. This will peak about 10 hour after the merger at an absolute optical  magnitude of $\sim -14$, corresponding to an apparent magnitude of 22.5 for a source at 200Mpc. This signal is beamed to within about $0.5$ rad. 
An isotropic signal at a comparable optical brightness could arise a few hours after the merger from the radioactive decay of the envelope component of the cocoon. 
Remarkably these could very well be the strongest and easiest to detect broad angle EM counterparts of gravitational radiation signals from these events.

To conclude we presented here a theoretical framework for calculating the emission from GRB cocoons as a function of the cocoon properties at the time of breakout. We then use a canonical model of such a cocoon to predict the expected broad-band signature. We stress again that there are numerous uncertainties in our approximations.
The most important one has to do with the amount of  mixing that takes place between the shocked jet and the shocked stellar matter. This is still needed to be explored in detail for various jet and stellar properties. Still, one component, the isotropic shocked stellar material signature is rather robust and apart from the dependence on the progenitor mass and radius and on the GRB jet luminosity and opening angle we don't see much leeway in this signal. 

We thank Ore Gottlieb and Richard Harrison for useful discussions. This research was supported by the I-Core center of excellence of the CHE-ISF. EN was was partially supported by an ERC starting grant (GRB/SN) and an ISF grant (1277/13). TP was partially supported  by an advanced ERC grant TREX.

\bibliographystyle{apj}

\end{document}